# Surface Alignment, Anchoring Transitions, Optical Properties, and Topological Defects in the Thermotropic Nematic Phase of an Organo-Siloxane Tetrapodes

Young-Ki Kim,[a] Bohdan Senyuk,[†a] Sung-Tae Shin,[b] Alexandra Kohlmeier,[c] Georg H. Mehl,[c] and Oleg D. Lavrentovich*[a]

[a] *Liquid Crystal Institute and Chemical Physics Interdisciplinary Program, Kent, OH 44242, USA;* *E-mail: olavrent@kent.edu*
[b] *Liquid Crystal Display Research and Development Center, Samsung Electronics Corporation, Korea*
[c] *Department of Chemistry, The University of Hull, Cottigham Road, Hull HU6 7RX, United Kingdom*
† *Present address: Department of Physics, University of Colorado, Boulder, Colorado 80309, USA*

*We perform optical, surface anchoring, and textural studies of an organo-siloxane "tetrapode" material in the broad temperature range of the nematic phase. The optical, structural, and topological features are compatible with the uniaxial nematic order rather than with the biaxial nematic order, in the entire nematic temperature range -25°C < T < 46°C studied. For homeotropic alignment, the material experiences surface anchoring transition, but the director can be realigned into an optically uniaxial texture by applying a sufficiently strong electric field. The topological features of textures in cylindrical capillaries, in spherical droplets and around colloidal inclusions are consistent with the uniaxial character of the long-range nematic order. In particular, we observe isolated surface point defect-boojums and bulk point defects-hedgehog that can exist only in the uniaxial nematic.*

## 1. Introduction

A uniaxial nematic ($N_u$) is a liquid crystal (LC) showing uniaxial anisotropy of physical properties with a single director $\hat{\mathbf{n}} \equiv -\hat{\mathbf{n}}$ being the symmetry axis and the optic axis.[1] Since the theoretical prediction in 1970,[2] there is a growing interest to the biaxial nematic ($N_b$) phase with three directors $\hat{\mathbf{n}} \equiv -\hat{\mathbf{n}}$, $\hat{\mathbf{m}} \equiv -\hat{\mathbf{m}}$, and $\hat{\mathbf{l}} \equiv -\hat{\mathbf{l}}$. The $N_b$ phase was first observed experimentally by Yu and Saupe[3] in a lyotropic mixture potassium laurate-decanol-water system. Although it is generally assumed that the biaxial phase is well documented in this case, there are reports that the biaxial optical features are only transient and when it is left intact, the samples eventually relax into the uniaxial state.[4,5] A more recent example of $N_b$ in a lyotropic LC has been presented in dispersions of board-like platelets.[6] The presence of $N_b$ is documented for thermotropic polymer LCs by solid state NMR spectroscopy, see Ref. 7 and references therein.

The low-molecular weight thermotropic version of $N_b$ is of a special interest, as it would allow one in principle to construct fast-switching displays and other electro-optic devices.[8-10] Design and synthesis of appropriate molecules have proven to be difficult. Simple rod-like mesogens explored so far do not yield a spontaneous biaxial order, although some of them do show a field-induced biaxiality with very fast (nanoseconds) electro-optic response.[11] Significant synthetic efforts thus extended to novel and more complex molecular architectures, such as cyclic mesogenic oligomer[12] to find the evidence of $N_b$ phase behaviour.[13] Very promising candidates for potential spontaneous $N_b$ behaviour were discussed for the so-called bent-core molecules.[14-22] However, re-examination of their properties lead to a conclusion that some of these materials are in fact uniaxial nematics and that their apparent biaxial appearance is caused by factors other than the true long-range biaxial orientational order.[23-31] One of these mimicking factors is a surface anchoring transition (reorientation of a uniaxial director $\hat{\mathbf{n}}$ at a bounding substrate) which results in optical biaxial appearance of the cell.[25,27,28] The second mechanism is rooted in the standard protocol of determining the phase diagram by cooling and heating the samples; it turns out that the flow and bidirectional director tilt during the thermal expansion/contraction of the material also lead to biaxial optical features.[30] Last but not least, formation of cybotactic smectic-C short-range clusters in a uniaxial nematic environment[31-34] of bent-core materials is one of the most often discussed mechanisms of how the material can acquire an appearance of the biaxial nematic in X-ray and other characterization techniques.[26,35-39] The smectic clustering might, on the other hand, facilitate a formation of the electric and magnetic field-induced biaxiality.[40-42] In this regard, it is of interest to note an extraordinary strong effect of the electric[43] and magnetic[44] fields on the phase diagram of the bent-core nematics, resulting in a shift of phase transitions by 4-12°C.

In this work, we explore an organo-siloxane material (Fig. 1) with a molecular structure that resembles a tetrapode.[45-52] Originally,[45] the phase sequence for a material with this structure was determined as $N_b$ (37°C) $N_u$ (47°C)I, where I stands for the isotropic phase. Later on, Figueirinhas et al.[46] presented the phase sequence of a mixture with a nematic deuterated probe as $T_g$(-30°C)$N_b$ (0°C) $N_u$ (47°C)I, here $T_g$ is the glass transition temperature, based on extensive solid state NMR investigations, with the results being in line with those reported in Ref. 7. The difference in the transition temperatures, compared to the pure sample[45] was attributed to the admixing of the probe, indicating that the stability of the $N_b$ phase is strongly affected by external stimuli. A combination of XRD studies and fast field cycling experiments suggest that a local $C_{2h}$ symmetry of the material in the nematic state, supporting the view that local clustering may be crucial for the explanation of the formation of a biaxial nematic phase.[49] Polineli et al.[52] present an additional evidence of a $N_u - N_b$ transition of the tetrapode material shown in Fig.1 at 37°C, as originally reported;[45] the authors find that though the transition is observed in conoscopic and optical tests, it is not easily identifiable in the elastic and electric measurements. Tallavaara et al.[50] report that $^{129}$Xe NMR studies indicate a phase transition at around $16°C - 18°C$ but it is not clear whether the low temperature nematic phase is uniaxial or biaxial. Finally, Cordoyiannis et al.[48] state that high-resolution adiabatic scanning calorimetry do not show a discernible $N_u - N_b$ transition.

In this work, we use electro-optical and optical microscopy (polarized light, polarization-sensitive fluorescence, conoscopy) techniques to explore the nature of the nematic phase of the tetrapode material shown in Fig.1. The samples represent (i) flat layers of thickness between $4\,\mu m$ to $50\,\mu m$ confined between two solid plates (ii) round capillaries of diameter $50\,\mu m$ and $150\,\mu m$; (iii) spherical or nearly spherical freely suspended droplets of diameter between $5\,\mu m$ and $20\,\mu m$. The observed electro-optical, surface and topological features of these samples show that the material has a uniaxial order in the whole nematic temperature range studied, $-25°C \leq T \leq 46°C$.

## 2. Material and Techniques

Figure 1 shows the chemical structure of the tetrapodic material. Four mesogens are connected to the siloxane core through four siloxane spacers. We confirmed the phase transition as $T_g\,(-27°C)\,N\,(46°C)\,I$ in the regime of cooling with the rate $0.1°C/min$; the transition temperatures agree well, within $1-3°C$, with the previous studies,[46, 49] if one does not discriminate between the two versions of the nematic (N) phase.

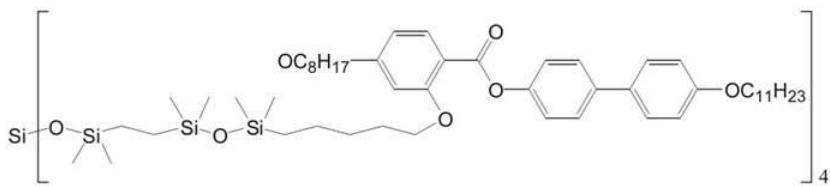

**Fig.1** Molecular structure of an organo-siloxane tetrapode.

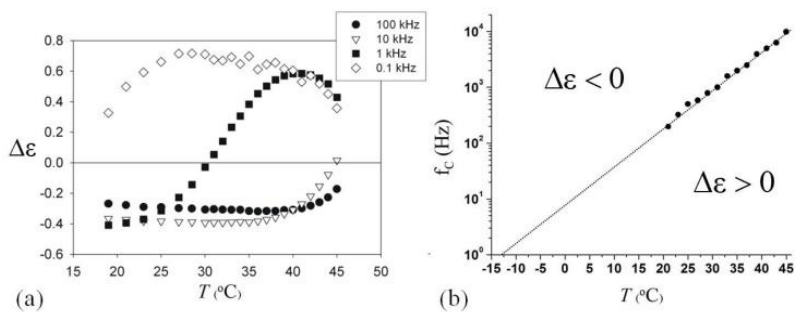

**Fig.2** (a) Dielectric anisotropy $\Delta\varepsilon$ and (b) crossover frequency $f_c$ as a function of $T$.

The flat cells were assembled from parallel glass plates with transparent indium tin oxide (ITO) electrodes. For planar alignment, the ITO glass substrates were spin coated with polyimide PI2555 (HD Microsystems); the polymer was rubbed unidirectionally. For homeotropic alignment (the director $\hat{\mathbf{n}}$ is perpendicular to bounding plates), the glass plates were treated with a weak solution of lecithin in hexane.

The cell thickness $d$ was set by spacers mixed with UV-curable glue NOA 65 (Norland Products, INC.) that was also used for sealing the cells. The actual cell thickness was measured by a light interference method. The material was filled into the cells in the isotropic phase. To prevent a possible memory effect, we performed most of the experiments for the temperatures above $-25°C$, which is slightly higher than the glass transition temperature $T_g = -27°C$. The temperature was controlled by a hot stage LTS350 with a controller TMS94 (both Linkam Instruments) with $0.01°C$ accuracy. A typical rate of temperature change was $\xi = \pm 0.1°C/min$ to minimize the effects caused by thermal expansion.[30, 53] Cooling was assisted by a circulation of liquid nitrogen.

In order to use the electric field as the means of director reorientation, we measured the dielectric properties of tetrapode with a precision LCR meter 4284A (Hewlett Packard) in the cell of thickness $d = 10\,\mu m$. We measured the dielectric permittivity for the directions parallel ($\varepsilon_\parallel$) and perpendicular ($\varepsilon_\perp$) to the director $\hat{\mathbf{n}}$ in homeotropic and planar cells, respectively, and thus determined the dielectric anisotropy $\Delta\varepsilon = \varepsilon_\parallel - \varepsilon_\perp$. We did not notice any biaxiality of the dielectric tensor when measuring the permittivity in the planar cells (which would have resulted in two different values of $\varepsilon_\perp$). To some extent, this result correlates with the previous studies by Merkel et al.[51] in which the biaxial dielectric anisotropy in the direction perpendicular to the main director was determined to be very weak, only about (-0.011). Such a small value of dielectric anisotropy does not provide a clear evidence of $N_b$ behaviour, as compared to other possible mechanisms, such as surface tilt, surface inhomogeneities or even surface anchoring transition that we observe in the tetrapode nematic. In the homeotropic cells, the director experiences a surface reorientation (an anchoring transition) when the temperature is lowered, as we discuss in a greater detail later. To reinforce the homeotropic alignment during the dielectric measurement of $\varepsilon_\parallel$, the cells were kept in the magnetic field of 1.4 T, directed normally to the cell. It allowed us to keep $\hat{\mathbf{n}}$ perpendicular to the glass plates in the range of temperatures $20°\text{C} < T < T_{NI}(46°\text{C})$, where $T_{NI}$ is the N-I transition temperature; the diamagnetic anisotropy of tetrapode material is positive. The dielectric anisotropy $\Delta\varepsilon$ changes its sign depending on temperature and frequency, Fig. 2a. Figure 2b shows the crossover frequency $f_c$ that separates the regions of different sign of $\Delta\varepsilon$, as the function of temperature being in line with those reported earlier.[52] The solid line is a tentative extrapolation of the dependency to lower temperatures. The condition $\Delta\varepsilon > 0$ can be used to distinguish the surface anchoring transition of a uniaxial $N_u$ phase from a hypothetical appearance of two optic axes if the material was a $N_b$ phase.

## 3. Results and Discussion

### 3.1 Homeotropic Alignment and Its Temperature Dependence

Optical tests of homeotropic cells offer a straightforward approach to determine whether the phase is $N_u$ or $N_b$ of orthorhombic symmetry provided there are no surface induced transitional effects. The homeotropic $N_u$ shows no birefringence for the orthoscopic transmission of light, regardless of its polarization. The cell is dark when viewed between two crossed polarizers. The homeotropic cell filled with the $N_b$ phase of orthorhombic symmetry, however, should show in-plane birefringence, $\Delta n_{xy} = n_y - n_x \neq 0$; its polarizing microscopy (PM) textures can be bright, depending on the orientation of the secondary directors $\hat{\mathbf{m}}$ and $\hat{\mathbf{l}}$ with respect to the crossed polarizers.

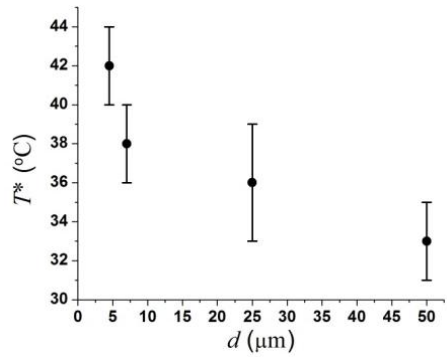

**Fig. 3** Plot of the anchoring transition temperature $T^*$ as a function of cell thickness.

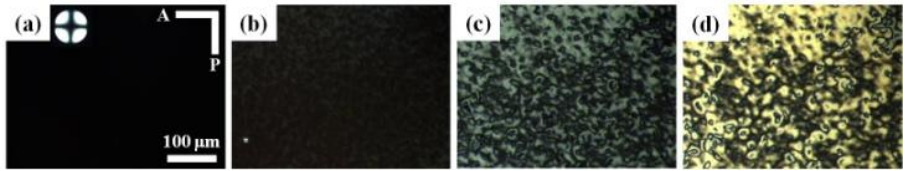

**Fig. 4** Transition from (a) dark uniaxial texture to (b-d) birefringent textures in the homeotropic cell ($d = 4.5$ μm) at (a) $T = 45°\text{C}$, (b) $40°\text{C} (= T^*)$, (c) $35°\text{C}$, and (d) $20°\text{C}$; inset in (a) is a conoscopic image of the cell.

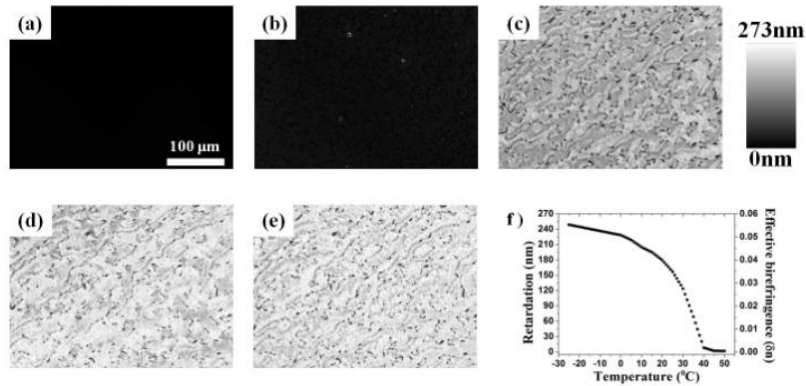

**Fig. 5** (a-d) Retardation maps and (f) local retardation as a function of temperature in a homeotropic cell ($d = 4.5$ μm) at (a) $T = 45°\text{C}$, (b) $40°\text{C} (= T^*)$, (c) $35°\text{C}$, (d) $30°\text{C}$, and (e) $20°\text{C}$.

At relatively high temperatures, above a critical temperature $T^*$, the tetrapode nematic in the homeotropic cells always show the standard homeotropic dark texture when viewed between two crossed polarizers, since the optic axis $\hat{\mathbf{n}}$ is parallel to the direction of polarized light propagation. The value of $T^*$ varies in a wide range $31°\text{C} \leq T^* \leq 44°\text{C}$, depending on the cell thickness $d$; $T^*$

decreases as $d$ increases (Fig. 3). Below $T*$, the dark homeotropic texture (Fig. 4a) becomes birefringent (Fig. 4b-d) with brightness increasing when the temperature decreases.

Using LC-Polscope (Abrio Imaging System), we mapped the optical retardance across the cells (Fig. 5a-e) and also measured the change of retardation $\Gamma$ (Fig. 5f) in a preselected location as the function of temperature for the homeotropic cell of thickness $d = 4.5\,\mu m$. The transition between the textures is reversible; the birefringent texture becomes dark when the temperature increases above $T*$. The very fact that the temperature $T*$ depends on the cell thickness (Fig. 3), suggests that the transition at $T*$ is not a bulk phase transition such as the $N_u$-to-$N_b$ transition, but could be a feature associated with surface effects. The data below support this conjecture.

There are two possible mechanisms for the transformation from the dark to the bright birefringent texture as the temperature changes: 1) the phase transition from $N_u$ to $N_b$ and 2) an anchoring transition, i.e. surface-mediated realignment of $\hat{n}$ in the $N_u$ phase[27, 28] (a combinations of both is also possible, of course). Anchoring transitions caused by temperature changes are well documented in various nematic LCs[54-56] and occur whenever there are two competing tendencies in setting the direction of surface orientation of $\hat{n}$ with different temperature dependencies. In the case of tetrapodes, these tendencies can be associated with the temperature evolution of molecular conformations, surface interactions and packing. As discussed in Ref. 41, strong electric and magnetic fields could affect the molecular packing and the symmetry of the mesophase made of complex molecules such as tetrapodes. What is important is that if $\hat{n}$ aligns parallel to the applied electric field, there is a simple way to discriminate between the two mechanisms by applying the electric field normal to the substrates without inducing any phase change.[25, 27] In the $N_u$ case, the main director $\hat{n}$ would be aligned along the field and the homeotropic cell would restore the dark PM texture, while $N_b$ would remain birefringent because the secondary directors $\hat{m}$ and $\hat{l}$ lead to the in-plane anisotropy and birefringence.

Figure 6 shows the change of PM texture (Fig. 6a-i), conoscopic texture (inset in Fig. 6a-i), and transmittance (Fig. 6j) by applying the vertical electric field (sinusoidal wave) with a frequency below $f_c$ in a homeotropic cell $(d = 6.9\,\mu m)$. To align the director $\hat{n}$ parallel to the field, the frequency $f$ should be below $\sim 10\,Hz$ at $T < -5°C$, Fig. 2b. Note that for low frequencies, the measured dielectric anisotropy $\Delta\varepsilon$ is influenced by finite electric conductivity; the condition $\Delta\varepsilon > 0$ means that the combined action of the dielectric and electric current torques leads to the alignment of $\hat{n}$ along the field. Another complication is that a high electric field might cause a dielectric breakdown.[57, 58] We indeed observed such a breakdown in our system for voltages higher than 200-250 V, which is close to or smaller than the voltage needed to realign the director into a homeotropic state when the temperature is very low, $T < -5°C$. To avoid dielectric breakdown, we performed the electric field experiments only in the range $-5°C < T < 46°C$.

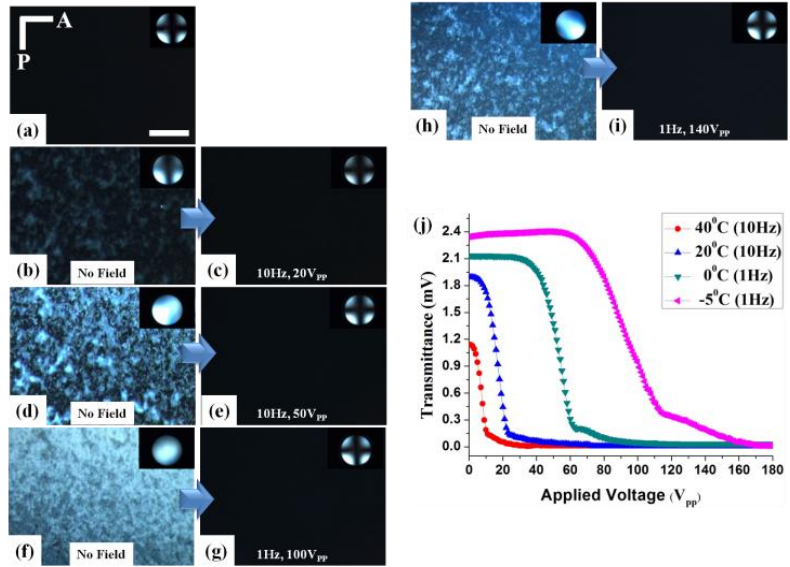

**Fig. 6** Change of PM texture and conoscopic pattern (a, b, d, f, h) without and (c, e, g, i) with a vertical field in a homeotropic cell ($d$ = 6.9 μm) at (a) $T$ = 45°C, (b, c) 40°C(= $T*$), (d, e) 20°C, (f, g) 0°C, and (h, i) -5°C; scale bar is 100 μm. (j) Transmittance vs. voltage curves for (b-i).

When $T > T*$, the homeotropic PM texture (Fig. 6a) corresponds to a symmetric Malthese cross (inset in Fig. 6a) in the conoscopic view, indicating that $\Delta n_{xy} = 0$ and $\hat{n}$ is aligned normal to the bounding plates. As the temperature decreases below $T*$, the PM texture becomes birefringent and the conoscopic cross becomes blurry (Fig. 6b,d,f,h and inset), indicating some misalignment. By applying a vertical field, however, we could always restore the dark PM texture and the symmetric Malthese cross (Fig. 6c,e,g,i and inset). Figure 6j shows that the transmittance of a He-Ne laser beam (wavelength $\lambda = 633\,nm$) passing through the nematic cell and two crossed polarizers always reduces to zero when the field is increased. This result demonstrates that the transformation of textures below $T*$ in the homeotropic cell is an anchoring transition rather than an $N_u - N_b$ phase transition. An alternative explanation would be that the electric field used in the experiments suppresses the biaxial order. Such a possibility seems unusual, as the electric field is directed along the principal director. In contrast to the results of Polineli et al.,[52] we do not observe an $N_u - N_b$ transition neither at 37°C nor at any other temperature in the homeotropic cells within the range $-5°C < T < 46°C$. We note that the surface tilt of the director becomes rather strong as the temperature decreases below the anchoring transition point $T*$ and thus at lower temperatures one needs higher electric fields to restore the uniaxial homeotropic configuration, Fig. 6.

Figure 7 presents a more detailed examination of the conoscopic patterns as the function of temperature (Fig. 7a) and field (Fig. 7b-d). The temperature induced modification of the conoscopic pattern is mostly in the shift of the center of the Malthese cross, which is consistent with the tilt in a $N_u$ phase rather than with the appearance of $N_b$ phase. The symmetric Malthese cross and the homeotropic uniaxial state of the cells are always restored at all the temperatures studied, if the applied electric field is of the proper frequency (to align $\hat{n}$ parallel to the field) and amplitude (to overcome the surface anchoring forces that do not favour a homeotropic state below $T^*$). In order to avoid a false biaxiality of the samples caused by thermal expansion,[30, 53] the rate of temperature change was kept low, $\pm 0.1°C/\min$.

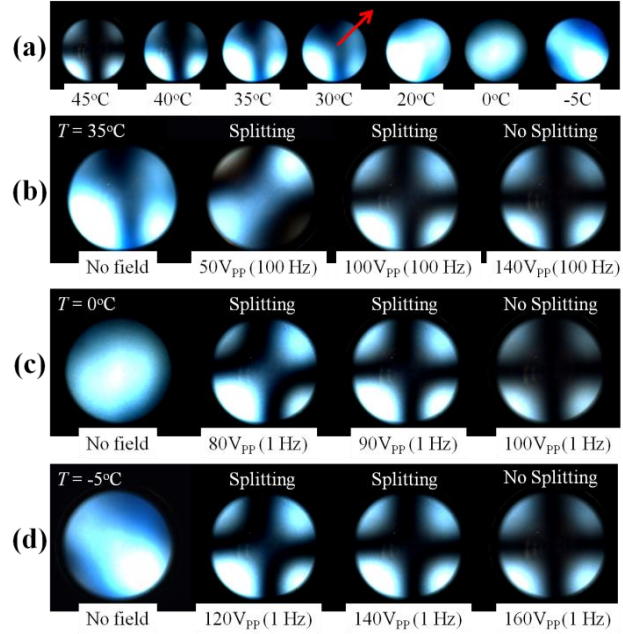

**Fig. 7** (a) Change of conoscopic texture as a function of temperature $T$ in a homeotropic cell; no electric field; (b-d) restoration of the Malthese cross of the uniaxial homeotropic state at $T < T^*$ by a vertical electric field ($d = 6.9$ μm) at (b) $T = 35°C$, (b, c) $0°C$, (d) $-5°C$. Red arrow in (a) indicates the shift direction of the Malthese cross.

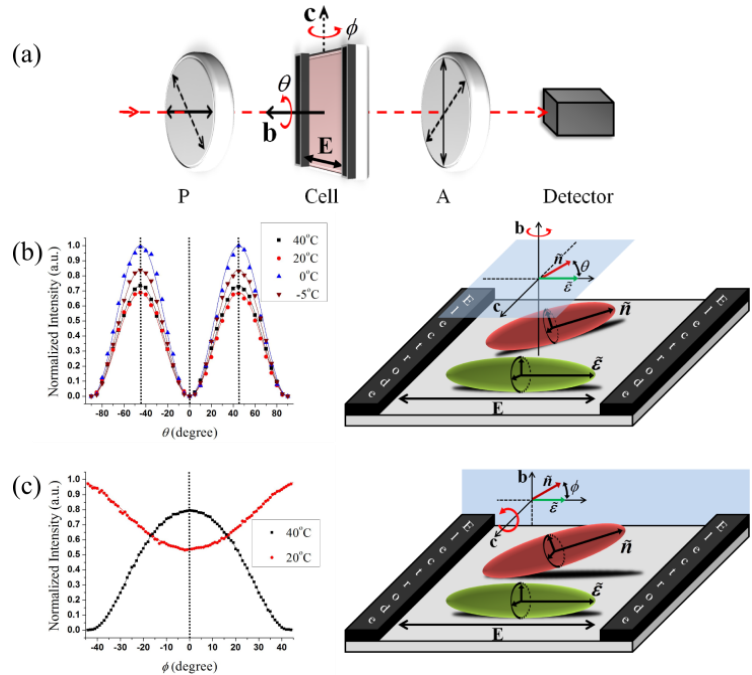

**Fig. 8** (a) Experimental setup to record the light transmittance as a function of cell rotation by $\theta$ around **b** axis and by $\phi$ around **c** axis, in the presence of in-plane field; The directions of polarizers for $\theta$ and $\phi$ rotation are indicated by solid and dotted lines, respectively. Intensity profile as a function of the angle (b) $\theta$ and (c) $\phi$ for different temperatures with an in-plane field in a planar cell ($d = 20$ μm); cartoons illustrate the verifiable geometries of monoclinic symmetry in which the low-frequency dielectric tensor and optical tensor have different directions.

The optical studies presented above present an argument against the existence of an $N_b$ phase with orthorhombic symmetry. As discussed by Tschierske and Photinos,[42] in the biaxial nematic phase $N_b$ of a lower monoclinic symmetry, the principal axes of the low-frequency dielectric tensor ($\tilde{\varepsilon}$) and the refractive index tensor ($\tilde{n}$) might not be parallel to each other. In that is the case, "a biaxial nematic phase with monoclinic symmetry can under special conditions appear optically isotropic due to the deviation of direction of the optical axis from the direction of the tilt axis".[42] In terms of the experimental situation described in Figs. 3-7, it means that the field-oriented state that appear dark ("isotropic") when viewed between crossed polarizers "from above" (with light propagating parallel to the applied electric field) might still be an $N_b$ phase of monoclinic symmetry. If this is the case, then optical observations with light propagating *perpendicularly* to the direction of the applied electric field would help to discriminate between such a monoclinic $N_b$ state and a normal uniaxial $N_u$ state. If the nematic is uniaxial, the direction of the applied electric field will be the axis of symmetry. If the nematic is biaxial of monoclinic symmetry, then the field direction would not be the symmetry direction of the optical response. The results of additional tests of a low-symmetry state are presented in Fig. 8 and discussed below.

To distinguish between a monoclinic biaxial and uniaxial states, we prepared planar cells ($d = 20$ μm) in which the electric field was applied in the plane parallel to the bounding glass plates. Both substrates are rubbed in an antiparallel fashion and the rubbing direction is tilted by 3° from the direction of field. We recorded the intensity of 633 nm light transmitted through the cell and a pair of crossed polarizers as the function of temperature and applied sinusoidal AC electric field: 2.92 ($f = 10$ Hz, $T = 40$°C), 4.12 (10 Hz, 20 °C), 7.28 (1 Hz, 0°C), 15.12 $V_{pp}$/μm (1 Hz, -5°C). The transmission was recorded as the function of two types of cell rotation. First, the cell was rotated by an angle $\phi$ around the axis "**b**" that is perpendicular to the plane of the cell (Fig. 8a,b, $T = 40, 20, 0, -5$°C). This experiment tests the potential tilt of the optic tensor with respect to the low-frequency dielectric tensor in the plane of cell, see the scheme in Fig. 8b. Second, the cell was rotated by an angle $\theta$ around the axis "**c**" that is perpendicular to the electric field but is parallel to the plane of cell (Fig. 8a,c, $T = 40, 20$°C). This experiment tests whether the optical ellipsoid is tilted away from the dielectric ellipsoid in the cross-section of the cell. In the second case, the experiments could be conducted only for temperatures above 20°C.

In both experiments, Fig. 8b,c, the intensity of transmitted light changes symmetrically around the values $\phi = \theta = 0$, thus indicating that the dielectric and refractive index principal axes are parallel to each other and to the direction of the applied electric field in the plane of the cell. Similarly to the results obtained for the homeotropic cells, Figs. 3-7, the data in Fig. 8 for the planar cells support the conclusion that the nematic phase in the range of temperatures studied is a uniaxial rather than biaxial (orthorhombic or monoclinic) nematic.

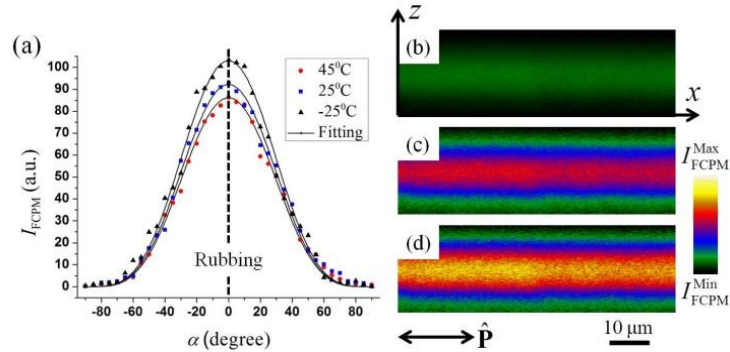

**Fig. 9** (a) FCPM intensity in a homogeneous planar cell of tetrapode material as a function of the angle $\alpha$. Vertical optical slices (*x-z* scan) of FCPM textures in the homeotropic cell ($d = 20$ μm) at (b) $T = 45$°C ($>T^*$), (c) 25 °C($<T^*$), and (d) -25°C.

Another clear evidence of the fact that the textural changes correspond to an anchoring transition in homeotropic cells of the $N_b$ state rather than to the appearance of $N_b$ state was obtained with a fluorescence confocal polarizing microscopy (FCPM). The tetrapode material was doped with a small amount ($0.01$ wt%) of a fluorescent dye *N,N'*-bis(2,5-di-tert-butylphenyl)-3,4,9,10-perylenedicarboximide (BTBP, Sigma-Aldrich). First, rubbed planar cells were prepared to establish the fluorescent dipole $\hat{\mathbf{d}}$ of BTBP dye. The fluorescent intensity $I_{FCPM}$ depends on the angle $\alpha$ between the polarization of probing beam $\hat{\mathbf{P}}$ and $\hat{\mathbf{d}}$ as $I_{FCPM} \propto \cos^4 \alpha$.[59, 60] Figure 9a shows $I_{FCPM}$ of the tetrapode nematic as a function of $\alpha$ at $T = 45$°C($>T^*$), $20$°C ($<T^*$), and $-25$°C. In the whole N temperature range, $\hat{\mathbf{d}}$ of BTBP is parallel to the main director $\hat{\mathbf{n}}$ of tetrapode nematic; $I_{FCPM}$ is max when $\hat{\mathbf{P}}$ is along the rubbing direction. Figure 9b-d show the vertical slices of FCPM texture in a homeotropic cell of thickness $d = 20$ μm; the cell was gently rubbed in anti-parallel fashion to reinforce unidirectional director tilting. In all cases, $\hat{\mathbf{P}}$ is parallel to the bounding plates. At $T = 45$°C ($>T^*$), the FCPM texture (Fig. 9b) has a low $I_{FCPM}$ which is corresponding to the dark PM texture indicating $\hat{\mathbf{n}}$ is aligned normal to the substrates. If the transformation of PM and conoscopic texture is not caused by the anchoring transition but by $N_u - N_b$ phase transition at $T < T^*$, $I_{FCPM}$ should be still low because of $\alpha = 90°$. However, at $T < T^*$, $I_{FCPM}$ increases which represents the tilting of $\hat{\mathbf{n}}$, Fig. 9c,d. Increase of $I_{FCPM}$ as $T$ decreases apparently demonstrates the tilting of director $\hat{\mathbf{n}}$. It is possible that $N_u - N_b$ phase transition and an anchoring transition occurs simultaneously but we demonstrated that there is no $N_u - N_b$ phase transition with the electro-optic measurement (Fig. 6, 7) and the observation of topological defects which is discussed below.

## 3.2. Topological Defects

### 3.2.1. Escape of Director in a Round Capillary

$N_u$ and $N_b$ have different sets of topological defects associated with a single director $\hat{n}$ in $N_u$ and three directors $\hat{n}$, $\hat{m}$, $\hat{l}$ in $N_b$.[61] Their features provide a useful phase/symmetry identification tool provided the investigation is carried not too close to the $N_u - N_b$ phase boundary (to avoid pre-transitional textural effects). In the bulk of $N_u$ phase, the topologically stable defects are point defects such as a "hedgehog" with a radial $\hat{n}(r)$ and linear defects-disclinations of strength $|m|=1/2$; here $m$ is the number of director turn by $2\pi$ when one circumnavigates the defect core once. Disclinations with $|m|=1$ are not topologically stable. Even if the configuration with $|m|=1$ is enforced by boundary conditions, for example, by confining $N_u$ into a round capillary, $\hat{n}$ would realign along the axis of capillary,[62, 63] a process called an "escape into the third dimension", Figure 10a.

In the $N_b$ phase, isolated point defects cannot be stable.[61] A radial configuration of one director in $N_b$ indicates that the two other directors are defined at a spherical surface and thus should form additional singularities emanating from the center of defect (bold red line in Figure 10b). The strength of disclination can be either ½ or 1. The disclination with $|m|=1$ are topologically stable with a singular core that cannot escape. Note that all these features of topological defects in $N_b$ should be relevant regardless of whether the symmetry of $N_b$ is orthorhombic or monoclinic.

Prior to filling LC, the round quartz capillary was treated with lecithin for homeotropic alignment. Figures 10c-g show the textural change as a function of temperature in the round capillary with $50\,\mu m$ (Fig. 10c,d) and $150\,\mu m$ (Fig. 10e-g) inner diameter (*D*). In the round capillaries with a homeotropic anchoring at bounding surface, the tetrapode samples show different textures, depending on *D* and *T*.

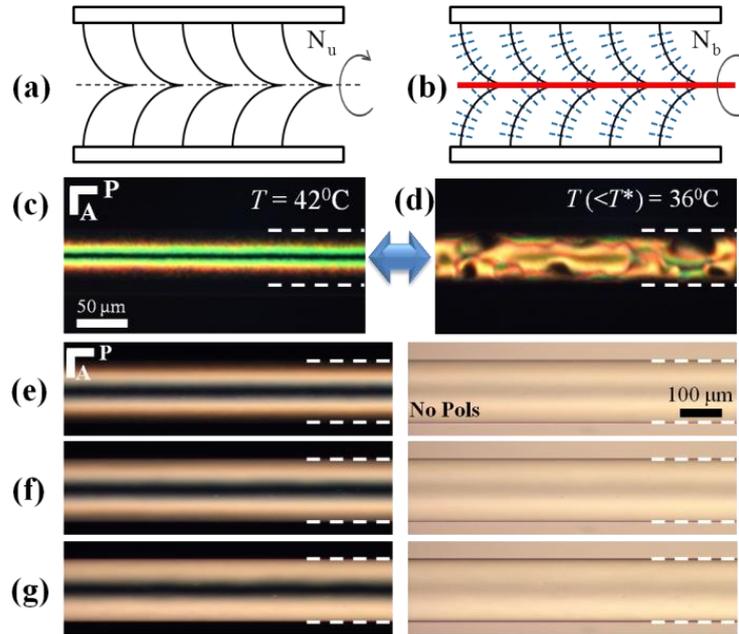

**Fig. 10** Escaped director configuration in a round capillary for (a) $N_u$ and (b) a hypothetical $N_b$; black solid, blue dashed, and red bold line represent the main director $\hat{n}$, secondary director $\hat{m}$, and disclination core, respectively. (c-g) PM textures of tetrapode nematic in the round capillaries; (c) capillary of diameter $D = 50$ μm shows an escaped configuration at $T = 42°C (>T^*)$ and a distorted texture due to an anchoring transition at (d) $T = 36°C (<T^*)$; large capillary of diameter $D = 150$ μm shows smooth texture of an escaped configuration with no biaxial features in the entire temperature range, (d) $T = 45°C$, (e) 25°C, and (f) -25 °C. White dashed lines in (c-g) indicate the inner wall boundaries of capillary

The narrow capillary, $D = 50$ μm, at $T > T^*$, shows the typical PM texture of the "escaped" director. When the capillary axis is parallel to one of the two crossed polarizers, one observes dark extinction bands in the center and near the capillary walls, Fig. 10c. The three dark bands are separated by two bright bands that correspond to the director orientation tilted with respect to the two polarizers. As the temperature decreases below $T^*$, the anchoring transition results in appearance of numerous defects and strongly deformed director, Fig. 10d.

In the wider capillary, $D = 150$ μm, the anchoring transition manifests itself in the tilt of the director at the inner wall of the capillary, thus the bright bands become wider, in the entire temperature range explored, $-25°C < T < 46°C$, Fig. 10e-g. The textures remain smooth and show no singularities, which is consistent with the escaped configuration characteristic for $N_u$; if the material were a biaxial $N_b$, this nonsingular configuration would be replaced with singularities which we do not observe.

### 3.2.2. Point Defects in Droplets Suspended in Isotropic Fluid

The surface anchoring determines the equilibrium state of director structure that in the case of spherical droplets must possess some number of topological defects, because of the geometrical theorems of Poincaré and Gauss.[54, 64] For the tangential anchoring, according to the Poincaré theorem, the spherical surface must contain point defects-boojums with the total strength $\sum_i m_i$ measured as the sum of two-dimensional topological charges $m_i$ equal 2. A typical realization in $N_u$ is the bipolar structure with two boojums at the poles, $m_1 = m_2 = 1$, Figure 11a. In $N_b$ phase, these boojums of strength $m=1$ cannot exist as isolated objects and they should represent the ends of singular disclinations of strength $m=1$ terminating at the surface; these disclinations do not vanish in the bulk (there is no "escape" mechanism for them), but they can split into disclinations of a smaller strength $m=1/2$; in that case, one would observe four exit points at the surface of the droplet, each of the strength $m=1/2$.[65] One possible realization is shown in Fig. 11b. Another possible $N_b$ structure is a single boojum of the strength $m=2$ that results when the disclinations shrink into a point at the surface, Fig. 11c. The value $m=2$ is the minimum value of the topological charge that an isolated surface point defect-boojum might have in $N_b$.

We produced the droplets of tetrapode material with tangential anchoring by dispersing it in glycerol (Sigma-Aldrich), Figure 11d-f. An important question is whether the limited solubility of glycerol in the liquid crystal (that is certainly possible, especially at elevated temperatures[66]), can dramatically change the phase diagram. We found that the temperature of melting of the droplets in glycerol is the same as for a bulk material (within a 0.2°C range, which is a typical variation from sample to sample), thus it is unlikely that glycerol alters the phase state and symmetry of the nematic material. The droplets with tangential anchoring show two isolated boojums in the entire studied temperature range $-25°C \leq T \leq 46°C$. No additional defects such as disclinations expected for the $N_b$ phase (bold line in Fig. 11b) are observed in these droplets.

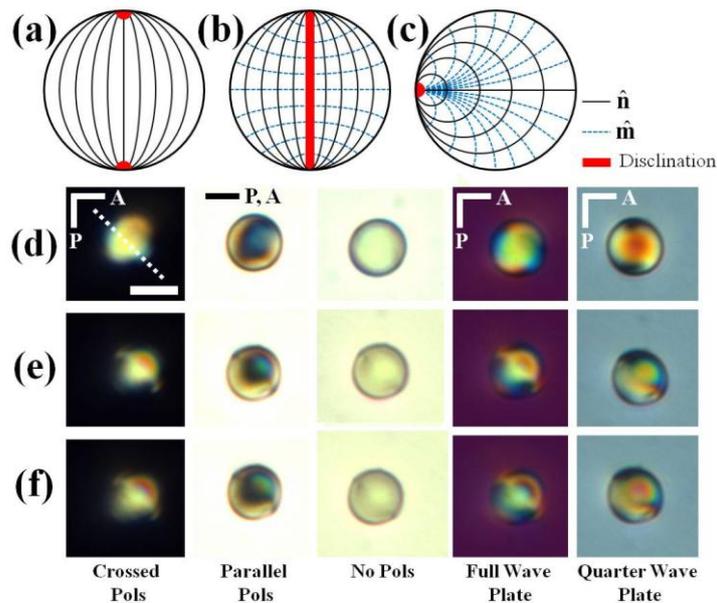

**Fig. 11** Director profile in the tetrapode droplet with tangential surface anchoring of $\hat{\mathbf{n}}$: (a) $N_u$ bipolar droplet with two point defects-boojums of strength $m_1 = m_2 = 1$ at poles; (b) hypothetical $N_b$ bipolar droplet with a singular disclination $m=1$ (red bold line) formed by secondary director; (c) hypothetical $N_b$ droplet with a single boojum of strength $m=2$. Textures of $N_u$ bipolar droplets of tetrapode material in a glycerol at (d) $T = 45°C$, (e) 25°C, and (f) -25°C. White dashed line indicates the droplet symmetry axis. Scale bar is 10 μm.

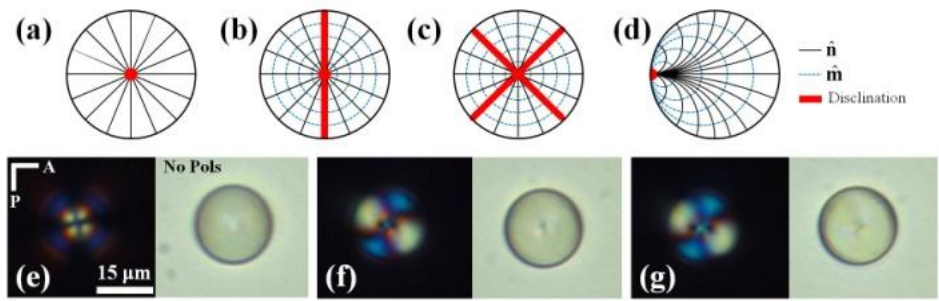

**Fig. 12** Director profile in tetrapode droplets with perpendicular surface anchoring of the director $\hat{\mathbf{n}}$: (a) $N_u$ radial droplet with a point defect hedgehog of strength $N=1$ at the center of droplet; Hypothetical $N_b$ droplets with normal anchoring for the principal director, (b) with singular disclination $m=1$ (red bold line) formed by secondary director, (c) with four disclinations of strength ½ each, and (d) a single surface point defect-boojum with $m=2$. The textures of tetrapode droplets in lecithin-glycerol mixture at (e) $T = 45°C$, (f) 25°C, and (g) 0°C.

Under the perpendicular anchoring condition, the equilibrium state of an $N_u$ spherical droplet is expected to contain a point defect of strength $N=1$ with a radial director field at the periphery, Fig. 12a. Its detailed core structure might be complex and contain a loop configuration, but the main point is that the defect is isolated, i.e., free of other defects attached to it. In $N_b$, however, the isolated point defects do not exist, as the appearance of the secondary directors implies that the point defect in the main director $\hat{\mathbf{n}}$ is connected to the surface by singular disclinations formed by the secondary directors, Fig. 12b,c. The disclination textures would most likely relax by minimizing the length of defects and forming a structure with one point defect-boojum at the surface, Fig. 12d; the latter is conceptually the same texture as in Fig.11c, with the interchange of the main and secondary directors.

To obtain the homeotropic anchoring, we added a small amount of a surfactant (lecithin) to glycerol. At high temperatures $T>T^*$, the droplets show crossed extinction brushes when observed between two crossed polarizers and an isolated point defect in the center, indicating the radial structure, Fig. 12a. As the temperature decreases below some critical value $T^*$, the crossed extinction lines start to deform. These changes of texture can be associated either with the surface anchoring transition or with the hedgehog losing its approximate radial configuration and acquiring deformations other than splay. The transition is reversible. Whatever the reason, the most important feature is that the hedgehog remains an isolated singularity that keeps its location at the center of droplet in the entire range $-25°C \leq T \leq 46°C$. We find no disclinations associated with the homeotropic droplets below $T^*$.

### 3.2.3. Point Defects at Colloidal Spheres in the Tetrapode Material

We also explored the behaviour of surface point defects - boojums and isolated point defects - hedgehog produced by spherical colloids dispersed in the tetrapode.

The spherical particles (diameter of 10 μm) of borosilicate glass were added into the tetrapode material and studied in homogeneous planar cells of thickness $d=20\mu m$. The glass yields a tangential anchoring and the texture shows a quadrupolar distortion of $\hat{\mathbf{n}}$ around the spheres with two point defects, boojums, at the poles (Fig. 13a-c and 13e). The axis connecting the two boojums of the same sphere is parallel to the overall director direction $\hat{\mathbf{n}}_0$ set by rubbing. Their existence follows from the same topological requirement $\sum_i m_i = 2$ as the existence of two boojums inside a $N_u$ drop. If the material surrounding the sphere is $N_b$, the isolated boojums cannot exist and should be accompanied by the singular disclinations, as shown, for example, in Fig. 13f. In our experiments, we find that the two boojums at the poles of the colloidal spheres have no disclinations attached, in the entire range $-25°C \leq T \leq 46°C$. It might be possible that the disclinations are simply invisible under the optical microscope. However, their presence should be manifested in the anisotropic forces of interaction of neighbouring spheres. In $N_u$, two tangentially anchored colloidal spheres attract each other in such a way that the line that connects their centres is tilted with respect to the overall director $\hat{\mathbf{n}}_0$.[67] In $N_b$ case, the disclinations should arrange the neighbouring spheres into a straight chain parallel to $\hat{\mathbf{n}}_0$, in order to minimize the length of disclinations. We do not observe this expected rearrangement, as the clusters of spheres show tilted arrangements expected for the $N_u$ environment, Fig. 13d.

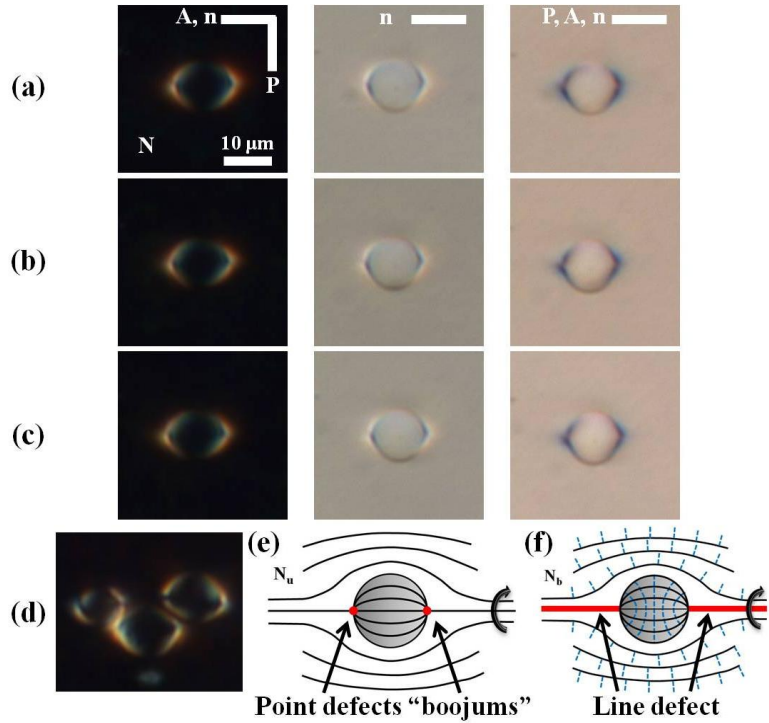

**Fig. 13** Point defects boojums $m_1 = m_2 = 1$ at the poles of colloidal spheres (diameter = 10 μm) in the planar cell ($d = 20$ μm) at (a) $T = 45°C$, (b) 25°C, and (c) -25°C: (d) chain of colloidal particles with point boojums at -25°C; (e) scheme of director configuration with point boojums in $N_u$ and (f) a hypothetical scheme of $N_b$ directors around the particle that preserves the same main director $\hat{\mathbf{n}}$ structure and thus needs to form new disclination defects in $N_b$ that are singular in the secondary director $\hat{\mathbf{m}}$. Black solid, blue dashed, and red bold line in (e) and (f) indicate main director $\hat{\mathbf{n}}$, secondary director $\hat{\mathbf{m}}$, and disclination core.

To explore the isolated point defects, so called "hyperbolic hedgehog" associated with homeotropically anchored director at the colloidal spheres,[67] the particles were treated with a surfactant octadecyltrichlorosilane (OTS). The particles were studied in a homogeneous planar cell $(d = 20\mu m)$. As shown in Fig. 14e, the spheres induce an isolated point defect that serves to compensate the topological charge of the sphere. In the entire temperature range $-25^\circ C \leq T \leq 46^\circ C$, the hedgehog retains its configuration, showing no appearance of the attached disclinations, Fig. 14a-c. The chain of particles was also stable, aligned along the direction of the elastic dipole, Fig. 14d, showing no rearrangements expected when the material acquires secondary directors.

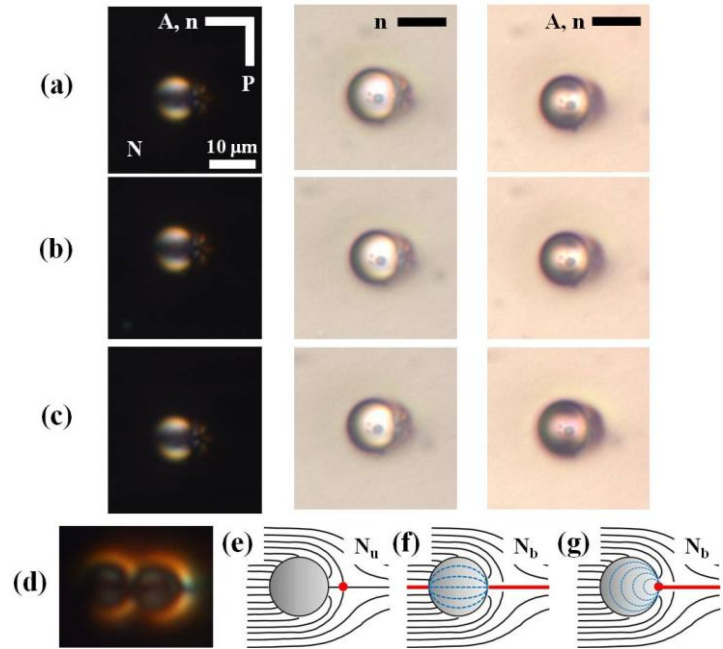

**Fig. 14** Isolated point defects "hedgehogs" formed next to each colloidal sphere (diameter = 10 μm) in the planar cell (d = 20 μm) at (a) $T = 45^\circ C$, (b) 25 $^\circ C$, and (c) -25 $^\circ C$: (d) chain of two colloidal particles with point defects at -25 $^\circ C$; (e) scheme of director configuration with a point defect "hedgehog" around the particles in $N_u$ and (f, g) hypothetical schemes of director field around the particle with line defects in $N_b$. Black solid, blue dashed, and red bold line indicate main director $\hat{n}$, secondary director $\hat{m}$, and disclination core.

## 4. Conclusions

We studied the optical, surface, and topological properties of nematic organo-siloxane tetrapodes to verify the existence of a biaxial nematic phase. In homeotropic cells, we observe a surface reorientation of the director below a certain temperature, $T < T^*$; $T^*$ varies from 31°C to 44°C depending on the cell thickness. We found that this transition is not necessarily associated with the appearance of the biaxial nematic phase, as one can restore the uniaxial optical character of the texture by applying a sufficiently strong electric field and aligning the liquid crystal director parallel to the field; the optical properties reveal a uniaxial character of the state. The possibility of a monoclinic biaxial order has been also verified in electro-optical response of the planar cells. The data, presented in Fig. 8, are consistent with the uniaxial order as opposed to a monoclinic biaxial order.

The external electric or magnetic field might in principle modify the packing of molecules in the nematic phase as evidenced by the field-induced shifts of melting points.[43, 44, 68-72] The field can also alter the symmetry of phase.[11, 73-78] For example, an electric field applied normally to the director $\hat{n}$ of the uniaxial nematic with a negative dielectric anisotropy can produce a biaxial modification of the order parameter.[11, 73-77] However, the field typically reduces the symmetry; we are not aware of the opposite effect of the field-induced transformation of the biaxial phase into a uniaxial phase. To avoid potential complications with the field-induced changes, we also tested the tetrapode nematic from the point of view of topological defects.

The studies of topological defects in various confinement geometries also support the conclusion of a uniaxial nematic order in the tetrapode material. In these studies, neither external electric nor magnetic fields are involved, and the liquid crystalline structures are determined by the surface anchoring conditions at the boundaries and by the intrinsic elasticity and symmetry of the nematic order. In round capillaries of a sufficiently large diameter, the director shows the "escape into the third dimension" texture that is stable in the entire nematic range of temperatures. In the cases of spherical confinement, such as nematic droplets dispersed in glycerine or spherical colloids dispersed in the tetrapode nematic, we observe isolated point defects-boojums and hedgehogs, which are consistent with the uniaxial type of ordering but not with the biaxial order. We thus conclude that in our experiments on optical, surface and topological features, the organo-siloxane tetrapode liquid crystal shows a uniaxial nematic phase behaviour in the range of temperatures $-25^\circ C < T < 46^\circ C$. This is at variance with the results based on other experimental techniques; a comprehensive picture of the molecular packing in the material is still outstanding.

## Acknowledgements

The work was supported by DOE grant DE-FG0206ER (dielectric response studies) and Samsung Electronics Corporation. ODL acknowledges the hospitality of Isaac Newton Institute, University of Cambridge, where part of the work was written. AK and GHM acknowledge funding through the EU project BIND.